\pdfoutput=1

\documentclass[11pt]{article}

\usepackage[preprint]{acl}

\usepackage{times}
\usepackage{latexsym}

\usepackage[T1]{fontenc}

\usepackage[utf8]{inputenc}

\usepackage{microtype}

\usepackage{inconsolata}

\usepackage{graphicx}
\usepackage{booktabs}
\usepackage{multirow} 
\usepackage{enumitem}

\title{Robustness Risk of Conversational Retrieval: Identifying and Mitigating Noise Sensitivity in Qwen3-Embedding Model}

\author{
 \textbf{Weishu Chen\textsuperscript{1}},
 \textbf{Zhouhui Hou\textsuperscript{2}},
 \textbf{Mingjie Zhan\textsuperscript{2}},
 \textbf{Zhicheng Zhao\textsuperscript{1,3}},
 \textbf{Fei Su\textsuperscript{1,3}},
\\
\\
 \textsuperscript{1}Beijing University of Posts and Telecommunications,
 \textsuperscript{2}SenseTime,\\
 \textsuperscript{3}Beijing Key Laboratory of Network System and Network Culture,\\
 \small{
   chenweishu@bupt.edu.cn
 } 
}

\begin{document}
\maketitle
\begin{abstract} We present an empirical study of embedding-based retrieval under realistic conversational settings, where queries are short, dialogue-like, and weakly specified, and retrieval corpora contain structured conversational artifacts. Focusing on Qwen3-embedding models, we identify a deployment-relevant robustness vulnerability: under conversational retrieval without query prompting, structured dialogue-style noise can become disproportionately retrievable and intrude into top-ranked results, despite being semantically uninformative. This failure mode emerges consistently across model scales, remains largely invisible under standard clean-query benchmarks, and is significantly more pronounced in Qwen3 than in earlier Qwen variants and other widely used dense retrieval baselines. We further show that lightweight query prompting qualitatively alters retrieval behavior, effectively suppressing noise intrusion and restoring ranking stability. Our findings highlight an underexplored robustness risk in conversational retrieval and underscore the importance of evaluation protocols that reflect the complexities of deployed systems. \end{abstract}

\section{Introduction}

Retrieval-augmented generation (RAG) and long-term memory mechanisms are increasingly deployed in conversational agents and assistant systems~\cite{rag_survey1, rag_survey2, rag_survey3}, where retrieval operates as part of an ongoing dialogue rather than as an isolated information-seeking step. In such settings, retrieval queries are often short, conversational, and weakly specified, reflecting intermediate dialogue states or memory requests rather than well-formed search intents. At the same time, retrieval corpora in deployed systems frequently contain heterogeneous artifacts, including system messages, dialogue logs, templates, and formatting residues produced during interaction. These characteristics place conversational retrieval in a regime that departs substantially from the clean-query, semantically coherent assumptions underlying standard embedding benchmarks~\cite{enevoldsen2025mmteb, muennighoff2023mteb_massivetextembedding}.

In this work, we identify and validate that this mismatch can expose a previously hidden robustness failure in modern embedding models. Focusing on Qwen3-embedding models~\cite{zhang2025qwen3embeddingadvancingtext}, a series of state-of-the-art models optimized for complex instruction following, we observe that under realistic conversational retrieval settings, structured dialogue-style noise such as greetings, polite buffers, or system-generated templates ,can become disproportionately retrievable and appear in at top-ranked positions, even though it is semantically uninformative. Notably, this effect emerges at low noise ratios and leads to severe ranking degradation, yet remains largely invisible under standard clean-query evaluations. Under the same evaluation protocol, this vulnerability is substantially more pronounced in Qwen3 embeddings than in earlier Qwen variants and several widely used dense retrieval baselines.

We find that a key factor amplifying this failure mode is the absence of query prompting~\cite{bge_embedding_cpack, e5_wang2024textembeddingsweaklysupervisedcontrastive, gritlm_muennighoff2024generative}. While prior work on instruction-tuned and prompt-based retrievers often characterizes prompting as a modest performance adjustment~\cite{zhang2025qwen3embeddingadvancingtext}, we reveal a novel finding that in conversational retrieval, prompting plays a qualitatively different role. Without prompting, Qwen3 embeddings exhibit extreme sensitivity to structured conversational noise; with even lightweight prompts, noise retrievability is largely suppressed and ranking stability is restored. This contrast indicates that prompting functions as a robustness gate that alters the retrieval regime itself, rather than as a weak optimization knob.

We empirically study this phenomenon across multiple datasets and conversational settings. Using controlled injection of non-adversarial, structured conversational noise, we systematically evaluate how noise intrudes into retrieval rankings and how prompting modulates this behavior. Our analysis spans multiple Qwen3 model scales and includes comparisons to closely related embedding models, as well as validation on a conversational memory benchmark that reflects realistic long-term usage.



Our contributions are threefold. (1) First, we identify a deployment-relevant robustness vulnerability in Qwen3-embedding models, where structured conversational noise can dominate retrieval results under realistic conversational conditions. (2) Second, we show that this failure mode is largely undetectable under standard clean-query benchmarks, highlighting a gap between benchmark evaluation and deployed behavior. (3) Third, we demonstrate that lightweight query prompting serves as an effective and practical mitigation, qualitatively suppressing noise retrievability rather than merely improving retrieval performance. 


\section{Conversational Noise Retrieval Setup}

We investigate embedding-based retrieval under conversational and memory-oriented settings, where retrieval is embedded in multi-turn interaction rather than being framed as an independent information-seeking task. Our goal is to evaluate how modern embedding models perform under realistic conversational conditions, especially when structured, dialogue-style noise naturally occurring in deployed systems.



\subsection{Structured Conversational Noise}
In conversational RAG, corpora often contain artifacts like system messages and dialogue logs, which depart from standard clean-query assumptions. We focus on structured, non-adversarial conversational noise that naturally arises from deployed dialogue systems and retrieval pipelines. Such noise is not designed to convey task-relevant information, but exhibits strong surface regularities that may interact with embedding similarity.

Concretely, we consider two broad categories of noise: (i) conversational fillers (e.g., 'I am ready to help', 'How can I assist you today?'), such as greetings, polite buffers, or assistant-style acknowledgments, and (ii) system- or format-level artifacts, including role prefixes, timestamps, system prompts, error logs, and serialized fragments (e.g., JSON- or XML-like patterns). All noise documents are task-agnostic and independent of retrieval queries. Detailed template instantiations are provided in the Appendix~\ref{appendix:noise_templates}.

\subsection{Noise Injection Protocol}
We construct the experimental corpus $\mathcal{D}_{total}$ by mixing noise documents $\mathcal{D}_{noise}$ into the original corpus $\mathcal{D}_{orig}$ at a specified ratio $\eta = |\mathcal{D}_{noise}| / (|\mathcal{D}_{orig}| + |\mathcal{D}_{noise}|)$. Noise documents are sampled independently of queries and are not optimized for adversarial effect. We evaluate a range of $\eta$ (typically $0\%$ to $15\%$) to assess retrieval stability, primarily using LongMemEval~\cite{longmemeval} as our testbed for ratio sweeps across different model scales.

\subsection{Evaluation Metrics}
We evaluate retrieval robustness using both ranking-sensitive and recall-based metrics. Our primary metric is NDCG~\cite{ndcg}, which captures ranking degradation when noise documents intrude into top-ranked positions. As structured noise primarily affects relative ordering rather than absolute retrieval success, NDCG provides a faithful signal of ranking instability under conversational noise.

We additionally report noise-specific indicators, such as the rank of the highest-ranked noise document or the presence of noise within the top-k results. These complementary metrics help disentangle overall retrieval performance from noise intrusion effects and support the primary analysis.

\begin{figure*}[ht]
    \centering
    \includegraphics[width=0.9\textwidth]{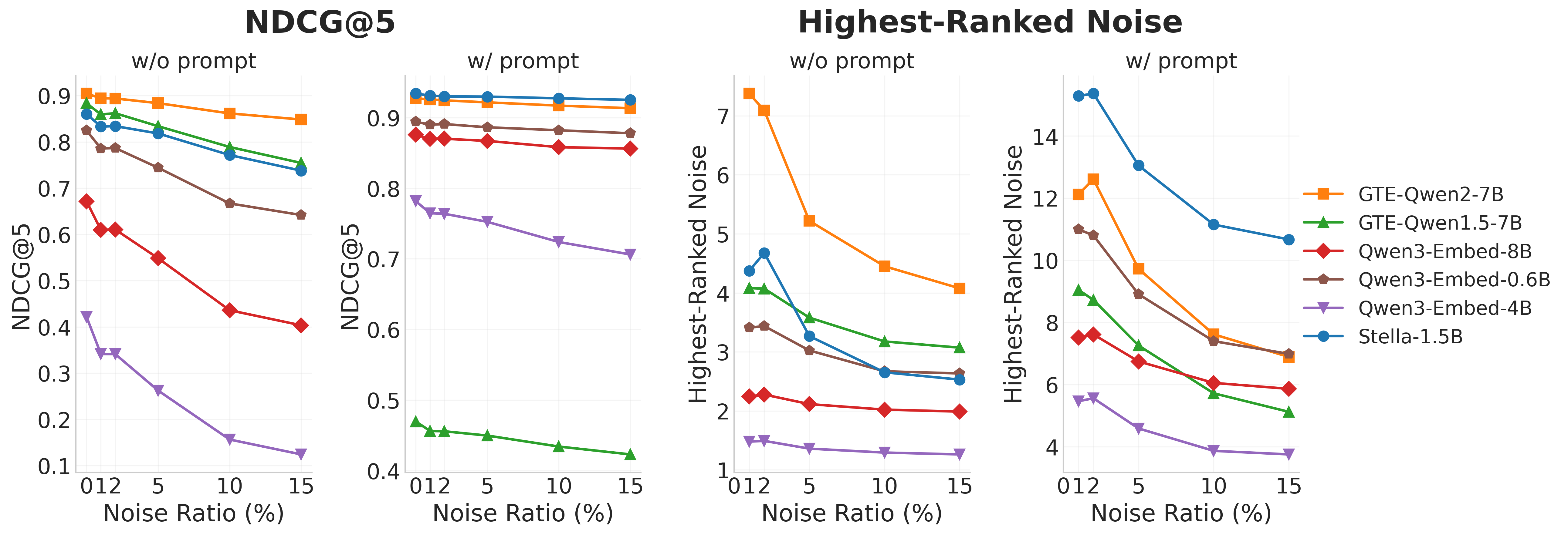}
    \caption{NDCG@5 and highest-ranked noise position versus noise ratio on LongMemEval (session-level). }
    \label{fig:curve_fig}
\end{figure*}

\section{Experiments}

This section reports experimental results on the impact of structured conversational noise in realistic retrieval settings. 

\subsection{Unique Fragility of Qwen3 Embeddings to Conversational Noise}
\label{subsec:main_result}

Figure~\ref{fig:curve_fig} presents retrieval robustness under increasing levels of structured conversational noise, measured by NDCG@5 and the rank of the highest-ranked noise document. Across all noise ratios, Qwen3-embedding models exhibit a qualitatively different behavior from other evaluated embeddings when query prompting is absent. Even at low noise ratios, Qwen3 embeddings suffer severe ranking degradation, with noise documents frequently appearing among the top-ranked results.

This vulnerability emerges consistently across Qwen3 model scales (0.6B, 4B, and 8B), indicating that the effect is not an artifact of a specific checkpoint or model size. For example, at a 1\% noise ratio without prompting, Qwen3-4B already incurs a substantial absolute drop in NDCG@5 relative to the no-noise condition, accompanied by noise intruding into the very top ranks. In contrast, other embedding models evaluated under the same protocol, including GTE~\cite{zhang-etal-2024-mgte1_5, gte2} variants and Stella~\cite{stella_zhang2025jasperstelladistillationsota}, exhibit markedly milder degradation, with both ranking quality and noise positions remaining comparatively stable. This contrast highlights Qwen3 as a clear outlier under conversational noise in the no-prompt setting.

Introducing query prompting fundamentally alters this behavior. With prompting enabled, Qwen3 embeddings largely recover their clean retrieval performance, and noise documents are consistently pushed to much lower ranks. Importantly, this change is not a gradual performance improvement but a qualitative shift in retrieval behavior: prompting effectively suppresses the noise-dominated retrieval pattern observed without prompting and restores ranking stability across noise ratios.

It is worth noted that, unlike the other retriever, GTE-Qwen1.5-7B is optimized for prompt-free retrieval. Consequently, introducing a prompt may cause a slight distribution shift that impairs its performance.

\begin{table}[t]
\centering
\small
\setlength{\tabcolsep}{3pt}
\begin{tabular}{lcc}
\toprule
\textbf{Noise Type} & \textbf{NDCG@5} & \textbf{Highest-Ranked Noise} \\
\midrule
No Noise                & 0.362 & --    \\
\midrule
Greeting                & 0.223 & 1.592 \\
Confirmation             & 0.243 & 1.876 \\
Apology                  & 0.245 & 1.976 \\
Suggestion               & 0.263 & 2.024 \\
\midrule
Assistant-style Greeting & 0.223 & 1.592 \\
User-style Greeting      & 0.224 & 1.616 \\
Descriptive Greeting     & 0.232 & 1.692 \\
\midrule
Error Logs               & 0.300 & 4.056 \\
System Prompts           & 0.234 & 1.830 \\
JSON Fragments           & 0.249 & 1.988 \\
Metadata Headers         & 0.275 & 2.753 \\
\bottomrule
\end{tabular}
\caption{Qwen3-Embedding 4B retrieval performance under diverse structured noise types at a fixed noise ratio (5\%) on LongMemEval (session-level). }
\label{tab:noise_type}
\end{table}

\subsection{Generality across Noise Types}

To verify that the observed vulnerability is not an artifact of a specific noise template or wording, we evaluate Qwen3-Embedding 4B under a diverse set of structured conversational and system-level noise types. Table~\ref{tab:noise_type} reports retrieval performance at a fixed noise (5\%) ratio on LongMemEval, including conversational fillers, stylistic variants, and structurally distinct system artifacts.

Across all evaluated noise categories, we observe substantial ranking degradation relative to the no-noise condition, with noise documents frequently appearing at top-ranked positions. This behavior is consistent across functionally different conversational templates (e.g., greetings, confirmations, apologies, and suggestions) as well as across stylistic variants of greeting noise, indicating that the effect is not driven by a particular phrasing, role framing, or conversational function.

Structural noise types exhibit some variation in severity, but the overall pattern remains stable. While error-log–like artifacts induce comparatively milder degradation, system prompts, serialized fragments, and metadata headers consistently disrupt ranking quality and introduce high-ranked noise. In short, these results demonstrate that the vulnerability identified in Section~3.1 holds broadly across diverse forms of structured conversational noise, and does note stem from a narrow or specific template choice.

\begin{table}[t]
\centering
\small
\setlength{\tabcolsep}{1.5pt}
\begin{tabular}{lcccccc}
\toprule
\textbf{Model} & \textbf{Pack} & \textbf{Prompt} &
\textbf{Clean} & \textbf{Noise} & \textbf{Noise Rank} \\
 &  &  & \textbf{NDCG@5} & \textbf{NDCG@5} &  \\
\midrule
Qwen3-0.6B & 0  & no   & 0.390 & 0.390 & 448 \\
           & 10 & no   & 0.564 & 0.563 & 45  \\
           & 10 & with & 0.572 & 0.572 & 60  \\
\midrule
Qwen3-4B   & 0  & no   & 0.351 & 0.351 & 384 \\
           & 10 & no   & 0.373 & 0.309 & 13  \\
           & 10 & with & 0.405 & 0.403 & 40  \\
\midrule
Qwen3-8B   & 0  & no   & 0.368 & 0.368 & 382 \\
           & 10 & no   & 0.437 & 0.391 & 20  \\
           & 10 & with & 0.493 & 0.492 & 53  \\
\bottomrule
\end{tabular}
\caption{Validation on LoCoMo. Clean (no-noise) and noise-injected retrieval performance (NDCG@5) of Qwen3 embeddings under different packing granularities. }
\label{tab:locomo}
\end{table}

\subsection{Effect of Memory Packing under Conversational Noise}

We further analyze the observed vulnerability under a common design in conversational retrieval systems in LoCoMo dataset~\cite{locomo_maharana2024evaluatinglongtermconversationalmemory}: aggregating multiple dialogue turns into coarser memory units. 


As shown in Table~\ref{tab:locomo}, coarser memory packing substantially improves retrieval performance in the clean (no-noise) setting across all Qwen3 model scales. This confirms that packing is an effective and practically motivated strategy for conversational memory retrieval, consistent with observations on LongMemEval.

However, under structured conversational noise, the same coarse-grained memory representation becomes markedly more vulnerable. In the no-prompt setting as packing size increases, noise documents intrude into very high ranks, accompanied by a clear degradation in NDCG@5 relative to the clean baseline. This indicates that structured conversational noise can effectively compete with aggregated memory units in the embedding space, amplifying the vulnerability identified earlier.

Introducing query prompting consistently mitigates this effect. With prompting enabled, noise intrusion is substantially reduced and retrieval performance under noise largely recovers to its clean counterpart, while preserving the benefits of memory packing. In summary, these results show that the vulnerability uncovered in Section~\ref{subsec:main_result} persists under realistic memory aggregation strategies commonly used in conversational retrieval systems.

\section{Discussion}

While a direct causal explanation remains unestablished, a plausible contributing factor to the observed vulnerability is the training paradigm of Qwen3-embedding models, which incorporates substantial amounts of synthetic data generated~\cite{zhang2025qwen3embeddingadvancingtext} by instruction-tuned large language models (Qwen3-32B~\cite{qwen3technicalreport}). Such data often exhibit strong conversational regularities, including greetings, polite buffers, and system-style templates. Under weakly specified conversational queries without prompting, these regularities may be preferentially activated in the embedding space, causing structured but semantically uninformative artifacts to become disproportionately retrievable. Lightweight query prompting appears to mitigate this effect by anchoring queries toward more task-oriented representations, thereby suppressing generic conversational priors. Importantly, this mitigation manifests as a qualitative shift in retrieval behavior rather than a gradual performance improvement, suggesting that prompting functions as a robustness gate in conversational retrieval settings.

\section{Conclusion}
In this work, we systematically examine the robustness risk of embedding-based retrieval under realistic conversational settings, focusing on structured, dialogue-style noise that naturally arises in deployed systems. Through extensive experiments, we show that Qwen-family embedding models, especially Qwen3, suffer from a scenario-dependent vulnerability, where such noise can intrude into top-ranked results despite being semantically uninformative. This effect is largely invisible under standard clean-query benchmarks but becomes pronounced in conversational and memory-oriented retrieval scenarios.

We further demonstrate that lightweight query prompting serves as an effective and practical mitigation, substantially suppressing noise retrievability across datasets, noise types, and memory configurations. Our findings highlight an underexplored robustness risk in advanced retrieval systems and underscore the importance of evaluating embedding models under deployment-relevant conditions. We anticipate that this study will inspire future efforts toward on robustness-aware evaluation and design of retrieval components for conversational and memory-augmented applications.

\section{Limitations}

\textbf{Diversity of Conversational Noise} While we evaluated a wide range of structured noise—from conversational fillers to system artifacts—the noise templates were generated based on common patterns in deployed systems. Actual production environments may contain more complex, nested, or model-specific artifacts (e.g., chain-of-thought residues) that were not covered in our controlled injection protocol.

\textbf{Scope of Model Vulnerability} Our findings primarily characterize the robustness failure in the Qwen3-embedding family. While our extended tests on other retrievers (e.g., Contriever~\cite{contriever}, E5~\cite{e5_wang2024textembeddingsweaklysupervisedcontrastive}) showed that they remain largely unaffected by such noise, we are currently unable to isolate the exact training samples or specific optimization objectives responsible for Qwen3's sensitivity. Due to the lack of detailed transparency regarding the proportions and specific templates of the LLM-generated synthetic data used during its development, it remains a challenge to definitively decouple the influence of training data distribution from other architectural or training factors.


\bibliography{custom}

\appendix


\section{Structured Conversational Noise Templates} 
\label{appendix:noise_templates}

In this section, we provide the full list of structured conversational and system-level noise templates used in the experiments described in Section 3.2.

\subsection{Conversational Fillers}

\paragraph{Greeting and Readiness}
\begin{itemize}[noitemsep]
    \item I'm here to help!
    \item How can I assist you today?
    \item I'm ready to assist you.
    \item I'm ready to assist you with anything you need.
    \item Hello! How may I help you today?
    \item Hi there! What can I do for you today?
    \item Hey! How can I help you today?
\end{itemize}

\paragraph{Understanding and Confirmation}
\begin{itemize}[noitemsep]
    \item I understand your question/concern.
    \item Got it.
    \item I see.
    \item Thanks for letting me know.
    \item I understand how frustrating that must be.
    \item That makes sense.
\end{itemize}

\paragraph{Apology and Empathy}
\begin{itemize}[noitemsep]
    \item I'm sorry to hear that.
    \item I apologize for the inconvenience.
    \item I'm really sorry you're experiencing this issue.
    \item My apologies for any confusion caused.
\end{itemize}

\paragraph{Suggestion and Solution Starters}
\begin{itemize}[noitemsep]
    \item Here's what you can try:
    \item I recommend this.
    \item You might want to try this.
    \item The best way to do this would be this.
    \item Let me walk you through the steps:
\end{itemize}

\subsection{Stylistic Variants of Greetings}

\paragraph{Assistant-style (Direct)}
\begin{itemize}[noitemsep]
    \item I'm here to help!
    \item How can I assist you today?
    \item I'm ready to assist you.
    \item I'm ready to assist you with anything you need.
    \item Hello! How may I help you today?
    \item Hi there! What can I do for you today?
    \item Hey! How can I help you today?
\end{itemize}

\paragraph{User-style (Inverted Perspective)}
\begin{itemize}[noitemsep]
    \item I'm here to get help!
    \item How can someone assist me today?
    \item I'm ready to be assisted.
    \item I'm ready to be assisted with anything I need.
    \item Hello! How may someone help me today?
    \item Hi there! What can someone do for me today?
    \item Hey! How can someone help me today?
\end{itemize}

\paragraph{Descriptive-style (Third-person)}
\begin{itemize}[noitemsep]
    \item The assistant is here to help.
    \item The assistant asks how it can assist the user today.
    \item The assistant is ready to assist the user.
    \item The assistant is ready to assist the user with anything needed.
    \item The assistant greets and offers to help the user today.
    \item The assistant asks what it can do for the user today.
    \item The assistant offers to help the user today.
\end{itemize}

\subsection{System and Format-level Artifacts}

\paragraph{Error Logs}
\begin{itemize}[noitemsep]
    \item {[ERROR]} Connection timeout
    \item {[WARNING]} Rate limit exceeded
    \item {[INFO]} Session initialized
    \item Exception: API call failed
    \item Retry attempt 3/5
    \item {[DEBUG]} Token count exceeded limit
    \item Status: 429 Too Many Requests
\end{itemize}

\paragraph{JSON Fragments}
\begin{itemize}[noitemsep]
    \item {\{"role": "assistant", "content": ""\}}
    \item {\{"timestamp": 1703750400, "turn\_id": 15\}}
    \item {\{"message\_id": "msg\_abc123", "type": "response"\}}
    \item {\{"session\_id": "sess\_xyz789", "status": "active"\} }
    \item {\{"model": "gpt-4", "temperature": 0.7\}}
    \item {\{"tokens": \{"input": 150, "output": 200\}\}}
    \item {[\{"role": "user"\}, \{"role": "assistant"\}]}
\end{itemize}

\paragraph{Metadata Headers}
\begin{itemize}[noitemsep]
    \item Session ID: sess\_xyz789
    \item Turn: 15/30
    \item Model: gpt-4-turbo
    \item Temperature: 0.7
    \item Max Tokens: 4096
    \item Context Window: 8192
    \item Response Time: 1.2s
    \item Token Count: 350
\end{itemize}

\begin{table}[h]
\centering
\small
\setlength{\tabcolsep}{1pt}
\begin{tabular}{lcccc}
\toprule
 &
\textbf{\shortstack{Clean\\Best Rank}} &
\textbf{\shortstack{Noise\\Best Rank}} &
\textbf{\shortstack{Highest-Ranked\\Noise}} &
\textbf{\shortstack{Prompt\\Best Rank}} \\
\midrule
Case 1 & 5 & 6 & 2 & 1 \\
Case 2 & 5 & 6 & 5 & 1 \\
\bottomrule
\end{tabular}
\caption{Qualitative case summary on LongMemEval. Structured conversational fillers intrude into top-ranked results under the no-prompt setting, while lightweight query prompting restores ranking stability.}
\label{tab:case_study_summary}
\end{table}

\section{Case Study: Qualitative Examples of Noise Intrusion}
\label{app:case_study}

To complement the aggregate metrics (e.g., NDCG and highest-ranked noise position), we present two qualitative examples from LongMemEval using Qwen3-embedding-8B that illustrate how structured conversational fillers intrude into top-ranked retrieval results under the \emph{no-prompt} setting, and how lightweight query prompting suppresses such intrusion. In both cases, the injected noise is semantically uninformative and independent of the query. The case results are summarized in Table~\ref{tab:case_study_summary}.

\subsection{Case 1: Noise ranks near the top and distracts retrieval}
\textbf{Query} \emph{``How many days ago did I launch my website when I signed a contract with my first client?''} \\
\textbf{Answer.} 19 days ago (20 days inclusive is also acceptable).

\noindent\textbf{Clean retrieval (no noise).} The gold document is retrieved at rank 5. The top results are dominated by long, off-topic dialogue logs; the correct memory appears within the top-5.\\
\textbf{Noise-injected retrieval (no prompt).} After injecting structured filler noise, a short template utterance (e.g., \emph{``I'm ready to assist you with anything you need.''}) intrudes into the top results and becomes the \emph{highest-ranked noise} at rank 2, while the gold document shifts slightly to rank 6. \\
\textbf{Noise-injected retrieval (with prompt).} With lightweight query prompting, the gold document moves to rank 1 and the injected filler noise no longer appears in the top-$k$ results.

\noindent\textbf{Ranks.} clean best rank = 5; noise best rank = 6; prompt best rank = 1; highest-ranked noise (no prompt) = 2.

\subsection{Case 2: Noise enters top-$k$ even when the answer is present}
\textbf{Query} \emph{``How many days before the 'Rack Fest' did I participate in the 'Turbocharged Tuesdays' event?''} \\
\textbf{Answer.} 4 days.

\noindent\textbf{Clean retrieval (no noise).} The gold document is retrieved at rank 5. The higher-ranked items are again long, off-topic conversational artifacts, but the correct memory remains within top-5. \\
\textbf{Noise-injected retrieval (no prompt).} With injected conversational filler noise, a short greeting-like template (e.g., \emph{``Hi there! What can I do for you today?''}) appears in the top results (highest-ranked noise at rank 5), and the gold document shifts to rank 6. Notably, the retrieval set contains \emph{both} the answer and the filler noise within top-$k$, indicating that the failure mode is not merely ``cannot retrieve the answer'' but rather a \emph{ranking instability} where semantically uninformative templates compete effectively in the embedding space. \\
\textbf{Noise-injected retrieval (with prompt).} With prompting, the gold document is retrieved at rank 1; the filler noise disappears from top-$k$.

\noindent\textbf{Ranks.} clean best rank = 5; noise best rank = 6; prompt best rank = 1; highest-ranked noise (no prompt) = 5.

\subsection{Observation}
Across both examples, the injected conversational fillers are short, highly regular, and query-independent, yet they intrude into top-ranked retrieval results under the \emph{no-prompt} setting. Lightweight query prompting consistently restores ranking stability: it promotes the gold memory to rank 1 and suppresses the retrievability of such generic templates. These qualitative cases align with the quantitative trend that prompting acts as a robustness gate in conversational retrieval.

\end{document}